# 'I Spend All My Energy Preparing': Balancing AI Automation and Agency for Self-Regulated Learning in SmartFlash

Hongming (Chip) Li, Salah Esmaeiligoujar, Nazanin Adham, Hai Li, and Rui Huang
University of Florida
{hli3, s.esmaeiligoujar, adhami.nazanin, li.ha, rui.huang}@ufl.edu

**Abstract:** Effective study strategies fail when preparatory tasks consume learning time. While AI educational tools demonstrate efficacy, understanding how they align with self-regulation needs in authentic study contexts remains limited. We conducted formative design research using an AI flashcard prototype, employing large language models to generate design hypotheses, which were validated through researcher walkthroughs and student sessions. Six students across disciplines completed sessions combining interviews and think-aloud tasks with their materials. Analysis revealed that students value automation for addressing the overwhelming preparation burden, yet require transparent, editable AI outputs to maintain cognitive ownership, which is essential for self-regulation. They conceptualized AI as a collaborative partner demanding verifiable reasoning rather than an autonomous agent. Metacognitive scaffolding was endorsed when clarifying study direction without constraining choice. Motivational features produced divergent responses. We derive design principles prioritizing editability and transparency, scaffolding metacognition without prescription, and accommodating motivational diversity. Findings identify conditions under which automation supports versus undermines metacognitive development in self-regulated learning.

## Introduction

Self-regulated learning is among the most significant predictors of academic success in higher education (Zimmerman & Schunk, 2011). However, empirical evidence consistently demonstrates that students struggle to implement effective self-regulation strategies, particularly in complex domains (Ainscough et al., 2018). The cognitive demands of preparing study materials exemplify this challenge, while others abandon proven study strategies entirely due to the prohibitive cognitive cost of implementation.

Recent advances in artificial intelligence offer potential solutions to these challenges through intelligent automation and adaptive support systems (Molenaar, 2023). Educational AI applications have successfully reduced mechanical tasks and provided personalized feedback (Jin et al., 2023), yet their integration into authentic study practices remains understudied. While efficacy studies proliferate, showing improved learning outcomes with AI-enhanced tools (Hussain et al., 2025; Agnes & Srinivasan, 2024), the field lacks systematic investigation of how these tools align with learners' needs and practices. More critically, we lack theoretical frameworks for understanding when automation supports versus undermines the development of self-regulation capabilities (Lan & Zhou, 2025).

Cognitive Load Theory provides a helpful lens for understanding these dynamics, distinguishing between the cognitive effort required for learning (intrinsic load), the effort dedicated to schema construction (germane load), and the effort wasted on inefficient processes (extraneous load) (Sweller, 2020). When preparatory tasks impose high extraneous load, they paradoxically prevent the deep processing they are meant to facilitate (Skulmowski & Xu, 2022). However, simply reducing this load through automation may create new problems if it diminishes learner agency or metacognitive engagement (Gupta et al., 2024). The challenge, therefore, is not merely technical but fundamentally about understanding how to design AI tools that reduce unnecessary cognitive burden while preserving the active engagement essential for self-regulated learning.

This tension between efficiency and agency emerges as a central design challenge in AI-supported education. While automation can eliminate tedious tasks, excessive automation risks creating passive consumers of pre-processed content rather than active, self-directed learners (Amershi et al., 2019). Understanding how learners perceive and interact with AI assistance in authentic study contexts becomes crucial for developing tools that enhance rather than replace self-regulation capabilities. However, current design methodologies often fail to capture the detailed interplay between cognitive, metacognitive, and affective factors that shape these interactions.

To address these gaps, we conducted a formative design study examining how higher education learners experience and interact with AI-supported study tools. We introduce a novel methodological approach that leverages large language models for hypothesis generation while maintaining rigorous human-centered validation. Through detailed analysis of learner experiences with an AI-powered flashcard prototype, we identify critical junctures where automation intersects with self-regulation, revealing both opportunities and risks in AI-supported learning design. Our research addresses three key questions: **RQ1)** What difficulties and needs do higher education learners report in authentic study practices? **RQ2)** Which cognitive and interactional points prove most problematic when using AI-supported study tools? **RQ3)** What design principles can guide the development of AI tools that effectively support self-regulated learning? Building on our human-centered analyses, we developed *SmartFlash*, an AI flashcard prototype that instantiates transparency, editability, and configurability to balance automation with agency.

**Figure 1**

*Core SmartFlash workflow illustrating the balance between automation and learner control. (Left) Learners upload or paste source materials and complete a brief AI-generated prior knowledge assessment before flashcards are generated. (Right) Generated flashcards are presented in an interactive study interface that supports navigation, answer reveal, and user-directed review.*

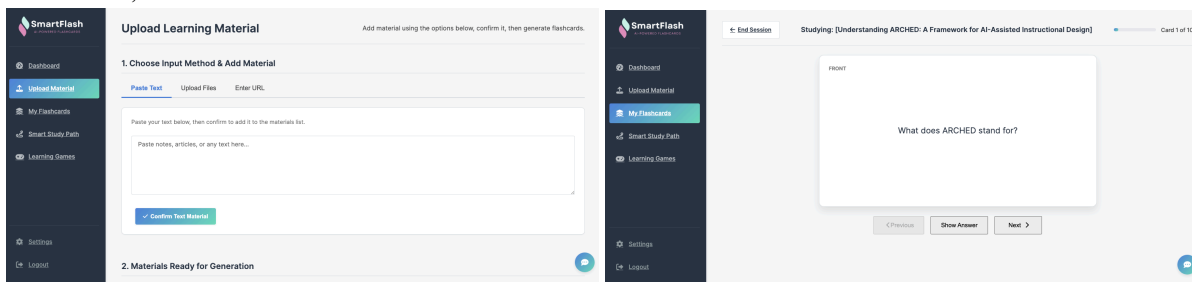

## Related work

### Theoretical framework

Three complementary theoretical frameworks inform our investigation of AI-supported self-regulated learning. Cognitive Load Theory (CLT) provides the foundational understanding of how mental resources are allocated during learning tasks (Sweller, 2020). CLT's tripartite model distinguishes intrinsic load inherent to the material, germane load supporting schema construction, and extraneous load arising from inefficient design or processes (Kalyuga, 2011). Research consistently demonstrates that excessive extraneous load impairs learning by consuming cognitive resources needed for deeper processing (Paas et al., 2004). In digital learning environments, this extraneous load often stems from interface complexity, poor information architecture, or laborious preparatory tasks (Skulmowski & Xu, 2022). Self-Regulated Learning (SRL) theory describes how successful learners actively manage their learning through cyclical planning, monitoring, and reflection (Zimmerman, 2002). Panadero's (2017) comprehensive review identifies six major SRL models, all of which emphasize metacognitive awareness as crucial for effective learning. However, developing these metacognitive skills proves challenging for many students, who struggle to accurately assess their understanding or identify knowledge gaps (Thiede & Dunlosky, 1994). This metacognitive challenge intensifies in autonomous online environments where external scaffolding is limited (Jansen et al., 2020). Recent work suggests that AI tools can externalize metacognitive processes through features like progress tracking and adaptive recommendations, potentially scaffolding SRL development (Molenaar et al., 2023). Human-Computer Interaction (HCI) principles, particularly the concept of co-adaptive systems, frame our understanding of learner-AI partnerships (Gallina et al., 2015). Unlike traditional educational software, which positions the system as instructor and the user as student, co-adaptive frameworks envision mutual adaptation in which humans and AI adjust their behaviors over time. This partnership model prioritizes user agency and control, recognizing that opaque automated systems can disempower learners and erode trust (Qin et al., 2020). Transparent, scrutable AI systems that explain their reasoning and allow user override have shown greater acceptance and effectiveness in educational contexts (Amershi et al., 2019).

### AI-enhanced learning tools and flashcard systems

Digital flashcard systems that leverage spaced-repetition algorithms have demonstrated robust effects on long-term retention across diverse domains (Chen, 2016). The testing effect, wherein retrieval practice enhances memory more than passive review, provides the psychological foundation for flashcard effectiveness (Schwieren et al., 2017). Recent innovations integrate AI to automate flashcard generation and personalize review schedules. Agnes and Srinivasan (2024) found that AI-generated mnemonic keywords significantly improved vocabulary retention compared to standard flashcards, suggesting that AI can enhance traditional study methods through intelligent augmentation. Similarly, Bachiri et al. (2023) demonstrated comparable quality between AI-generated and human-created flashcards in MOOC contexts, validating the potential for AI to reduce preparation burden without sacrificing educational quality. However, these studies primarily focus on learning outcomes rather than the design process or user experience. Odeleye and Dallas (2024) begin to address this gap by examining how personalized AI feedback affects cognitive load and engagement, finding that adaptive features must carefully balance challenge and support to maintain optimal cognitive load. Their work suggests that AI personalization succeeds when it responds to individual differences in prior knowledge and learning preferences. Nevertheless, they provide limited insight into how learners perceive and interact with these adaptive features during authentic study sessions.

Perhaps most critically, the literature lacks a detailed investigation of the automation-agency tension in AI-supported learning. While some researchers advocate maximum automation to minimize cognitive load (Hussain et al., 2025), others warn that excessive automation may undermine metacognitive development and learner autonomy (Gupta et al., 2024). Jin et al.'s (2023) systematic review notes this unresolved tension but offers limited empirical guidance for navigating it. Understanding how learners experience and negotiate this balance in practice remains essential for developing AI tools that enhance rather than replace self-regulated learning capabilities.

## Methods

### Participants, setting, and research design

Six participants (ages 22-31) were recruited from a large R1 public university in the southeastern United States through purposive sampling. The sample included graduate and senior undergraduate students from mathematics (P1), medicine (P2), educational technology (P3, P4), biology (P5), and engineering (P6). This disciplinary diversity captured varied perspectives on study practices in demanding academic contexts. This study was determined to be exempt by the Institutional Review Board (IRB) at the University of Florida (Protocol No. ET00048287). All sessions were conducted on *SmartFlash*. We developed a three-phase formative research process integrating AI-assisted hypothesis generation with human-centered validation. First, we prompted GPT-5 with descriptions of the target population and tool function to generate testable hypotheses about user needs, pain points, and behavioral patterns. These outputs formed a structured framework of assumptions requiring empirical validation. Second, two researchers independently conducted cognitive walkthroughs (Polson et al., 1992) of the prototype's core workflow, creating and reviewing flashcards from source texts, identifying potential usability issues and interaction breakdowns. Third, using the high-fidelity prototype, each participant completed a 60-minute recorded session combining semi-structured interviews about authentic study practices with think-aloud protocol tasks (Ericsson & Simon, 1998). Interview questions explored current study challenges before prototype exposure to avoid priming effects. Think-aloud tasks required participants to create flashcards from their own study materials while verbalizing thoughts, expectations, and reactions.

### Data analysis

Transcripts and session recordings underwent reflexive thematic analysis (Braun & Clarke, 2019). After initial familiarization, researchers conducted line-by-line coding to identify segments related to user needs, behaviors, and tool interactions. Codes were iteratively refined through constant comparison between interview statements and observed behaviors. Emerging themes were validated against the whole dataset and compared with AI-generated hypotheses to assess congruence, divergence, or supplementation. Analysis remained fully human-conducted, with AI outputs serving only as initial scaffolding.

## Results

Analysis revealed four themes characterizing the cognitive, metacognitive, and affective dimensions of learner experiences with AI-supported study tools.

### Theme 1: Preparatory burden as barrier

Participants consistently identified material preparation as the primary obstacle to effective studying. A medical student (P2) reported: "I had an exam and a four-month period for preparing for it, and three months just was gone by only preparing the flashcards... I didn't even get to review them all." A biology student (P5) stated: "I spend all my energy preparing instead of learning." The mathematics student (P1) described creating flashcards as cognitively overwhelming, leading to complete task abandonment despite knowing the strategy's effectiveness. During prototype interaction, automation of this process elicited immediate positive responses, with participants characterizing it as a "life-saver" (P5) and "really helpful" (P4). The AI hypothesis generation correctly identified time-consuming preparation as a pain point, though it underestimated the severity of impact on learning behaviors.

### Theme 2: Metacognitive uncertainty

Participants expressed persistent uncertainty about study direction and progress monitoring. P3 described being unable "to identify key points," leaving him "confused" about exam preparation. P5 linked this uncertainty to procrastination: "Without knowing what to do next, I get overwhelmed and avoid starting." During think-aloud sessions, metacognitive scaffolding features received strong endorsement. P1 found the "Smart Study Path" feature "absolutely useful," while P5 valued "Suggested Next Step" because "I struggle with knowing what to do next." While AI hypotheses identified "difficulty prioritizing," they missed the affective consequences—anxiety, overwhelm, and procrastination, that human analysis revealed as central to the experience.

### Theme 3: Agency and control requirements

Participants expressed sophisticated expectations for human-AI partnership, wanting assistance without sacrificing control. P6 stated: "I wouldn't just trust it immediately. I'd want to check if what it generates is accurate." P1 questioned whether AI "really gets what I need to focus on." This need for agency manifested most strongly in response to non-editable auto-generated content. P1 reported: "I wanted to edit, but it doesn't allow it. This makes me anxious." The inability to modify or delete generated flashcards emerged as the most frequent usability complaint across sessions. AI-generated hypotheses emphasized automation benefits but failed to anticipate this fundamental requirement for user control and verification.

**Figure 2**
*Gamification interface illustrating motivational scaffolding features. SmartFlash includes competitive and time-based study modes, as well as leaderboard and achievement tracking.*

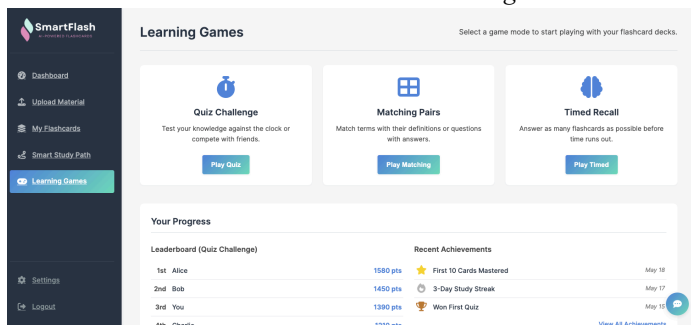

### Theme 4: Individual variability in motivational response

Gamification features elicited contradictory responses across participants. P3 viewed leaderboards positively: "To compare with my classmates, I think this design is very good. It will make me more eager." Conversely, P1 reported: "Higher score doesn't mean anything to me... It makes me very anxious. It demotivates me from playing the games." P5 requested alternative motivational mechanisms, like streak counters rather than social comparison. These divergent responses indicate that motivational features cannot assume universal appeal. The AI hypothesis generation did not predict this variability or the potential for features to simultaneously motivate and demotivate different users.

## Discussion

Our findings reveal a paradox: automation that reduces the preparatory burden can threaten the learner agency essential for self-regulation. When this preparatory burden consumes the majority of study time, as participants reported, we observe a systemic failure where scaffolding displaces learning. The medical student who spent three months preparing flashcards without reviewing them exemplifies how the preparatory burden transforms from support to barrier. The pervasive metacognitive uncertainty reported by participants illuminates a gap between knowing study strategies and when to deploy them. This extends Thiede and Dunlosky's (1994) work by revealing how anxiety and overwhelm create negative feedback loops that further impair metacognitive judgment.

While AI scaffolding through features like "Smart Study Path" can externalize guidance, participants valued recommendations and explanatory rationales. This preference for transparency suggests that adequate AI support helps develop rather than bypass metacognitive awareness. Participants conceptualized AI as a collaborative partner whose suggestions require verification and outputs demand curation. The frustration with non-editable auto-generated content reveals that editing serves deeper functions than error correction. It represents cognitive ownership, the process through which external information transforms into personal knowledge structures. This finding aligns with co-adaptation principles (Gallina et al., 2015) but highlights the pedagogical importance of user modification. Learners lose crucial mechanisms for active engagement and knowledge construction when AI outputs cannot be edited. The divergent reactions to gamification features reveal fundamental differences in motivational orientation. Competition that energized one participant induced anxiety severe enough to prevent engagement in another. These differences reflect distinct goal orientations requiring different support structures (Wolters & Taylor, 2012), challenging universal gamification approaches. This variability suggests that personalization in educational AI extends beyond content to encompass the entire interaction paradigm.

Our AI-seeded formative workflow provides methodological insights beyond specific findings. The LLM identified functional challenges, such as time consumption, but missed affective and agentive dimensions central to the user experience. This pattern highlights current AI's strength in recognizing surface features, while underscoring its inability to grasp the subtleties of lived experience. The divergence validates human-centered research as providing a grounded understanding of learning phenomenology that AI cannot capture.

These findings suggest effective educational AI requires navigating rather than resolving the tension between support and autonomy. The goal isn't maximum automation or complete learner control, but dynamic balance emerging through interaction. As AI capabilities expand, distinguishing what can from what should be automated becomes critical. Our data suggest that this distinction depends not on technical feasibility but on whether automation preserves active engagement for the development of self-regulated learners.

With six participants and single-session observations, depth is prioritized over generalizability and long-term outcomes. Future longitudinal research should examine whether prolonged AI support enhances or atrophies self-regulation capabilities using validated SRL instruments (Rivers et al., 2020). Studies with larger, more diverse samples could identify patterns in how different learner populations navigate the automation-agency tension. The field urgently needs frameworks for determining which aspects of learning should remain human activities regardless of technological capability. Research should investigate threshold effects in cognitive load interventions and develop methods for accommodating individual differences in motivational orientation within AI systems. As educational AI becomes more widespread, understanding these dynamics is essential for fostering capable, autonomous learners, rather than users who are overly reliant on automated support.